\documentclass{mn2e}
\usepackage{graphicx}

\title{Comptonization in the vicinity of black hole horizon}

\author[A. Nied\'zwiecki]
{Andrzej Nied\'zwiecki,\\
\L \'od\'z University, Department of Physics, 
Pomorska 149/153, 90-236 \L \'od\'z, Poland\\
}

\date{29 September 2004}

\pagerange{\pageref{firstpage}--\pageref{lastpage}}
\pubyear{2004}
\begin{document}
\maketitle
\label{firstpage}

\begin{abstract}

Using a Monte Carlo method, we derive spectra arising from
Comptonization taking place close to a Kerr black hole. We consider a
model consisting of a hot thermal corona Comptonizing seed photons
emitted by a cold accretion disc. We find that general relativistic
effects are crucial for the emerging spectra in models which involve
significant contribution of radiation produced in the black hole
ergosphere. Due to this contribution, spectra of hard X-ray emission
produced in the vicinity of a rapidly rotating black hole strongly
depend on the inclination of the line of sight, with larger
inclinations corresponding to harder spectra. Remarkably, such
anisotropy could be responsible for properties of the X-ray spectra of
Seyfert galaxies, which appear to be intrinsically harder in type 2
objects than in type 1, as reported recently.

\end{abstract}


\section{Introduction}
A large fraction of the energy liberated by accretion onto black holes
is released in an optically thin plasma, which produces hard X-rays by
Compton upscattering of soft photons. In some cases this hard X-ray
source must be located in the immediate vicinity of a black hole, as
indicated by Fe K$\alpha$ emission line profiles revealed by {\it
Chandra} and {\it XMM-Newton} (e.g.\ Wilms et al.\ 2001; Miller et
al.\ 2002). If those lines are produced at inner accretion discs,
their steep radial emissivities require illumination by hard X-rays to
be centrally concentrated within a few gravitational radii, which in
turn implies that the X-rays are generated close to the black hole
horizon. In such cases the X-ray spectra should be affected by Doppler
and gravitational energy shifts as well as ray bending effects. The
influence of strong gravity on Compton scattering expresses itself in
an extreme way in the ergosphere, where scatterings can extract
rotational energy of a black hole (Piran \& Shaham 1977a).

Despite the expected importance of these effects, there has been
little work done to investigate their impact on Comptonization spectra
emerging from black hole accretion flows. Piran \& Shaham (1977b)
considered production of $\gamma$-rays in the black hole ergosphere by
Compton upscattering of X-ray photons. However, their results do not
have a form suitable to compare them  with current observations of
black hole systems.  We find also that their conclusions are partially
incorrect.  More recently, Comptonization  in advection-dominated
accretion flows around Kerr black holes has been analyzed  by
Kurpiewski \& Jaroszy\'nski (1999) and Manmoto (2000).  However,
emission from such flows is dominated by radiation produced in regions
rather distant from a black hole, where effects of strong gravity are
not significant, as we find in this paper.  Laurent \& Titarchuk
(1999) derived spectra of Comptonization on a relatively cold plasma
moving with relativistic velocities close to a black hole. However,
these bulk  motion Comptonization models have assumed so far only the
Schwarzschild metric, omitting processes taking place in the
ergosphere, where strong gravity effects are most prominent.

In this paper we study formation of Comptonization spectra using a
fully general relativistic description of photon transfer and Compton
scatterings in a plasma located close to a Kerr black hole.  For our
investigation we choose a model in which seed photons are emitted by a
cold disc, while Comptonization takes place in a hot thermal corona.
We derive the cold disc emission and the Comptonization spectra by
following individual photon trajectories. We use a Monte Carlo method
to simulate both the thermal emission of the disc and Compton
scatterings.

We take into account two geometries of a disc-corona system  which are
popular in modeling the central part of black hole accretion flows.
Firstly, we consider a model involving inner corona replacing a cold
disc in the central region.  This geometry is similar to solutions
with a hot optically thin flow  existing in the inner part of a cold
disc (e.g.\ Shapiro, Lightman \& Eardley 1976;  Narayan \& Yi 1995;
Esin, McClintock \& Narayan 1997), although we assume a much lower
distance of transition between the disc and the inner corona than in
standard versions of this solution. Secondly, we consider a corona
surrounding a cold disc, corresponding to models of a hot dissipative
corona formed on top of the disc surface (e.g.\ Liang \& Price 1977;
Haardt \& Maraschi 1991).

In Section 2 we describe details of our model. In Section 3 we analyze
relevant relativistic effects. Examples of our model spectra are shown
in Section 4. We discuss our results in Section 5 and summarize our
conclusions in Section 6.

\section{Model description}

\subsection{Geometry and model parameters}

We consider a Kerr black hole  accreting matter at a mass accretion
rate $\dot M$. The black hole is characterized by its mass, $M$, and
angular momentum, $J$. We use the Boyer-Lindquist (BL) coordinate
system $x^i = (t,R,\theta,\phi)$. The metric tensor components,
$g_{ij}$, are given, e.g., in Bardeen, Press \& Teukolsky (1972). We
also make use of locally non-rotating (LNR) frames (Bardeen et al.\ 1972),
which are particularly convenient for studying processes in the Kerr
geometry.

The following dimensionless parameters are used in the paper
\[
r = {R \over R_{\rm g}},~~~\hat t = {ct \over R_{\rm g}}, 
~~~{\dot m} = { {\dot M} \over {\dot M}_{\rm Edd}}, 
~~~a = {J \over c R_{\rm g} M},
~~~\Omega = {{\rm d} \phi \over {\rm d} \hat t},
\]
where ${\dot M}_{\rm Edd} = 4 \pi G M m_{\rm p} / (\sigma_{\rm T} c)$
is the Eddington accretion rate and $R_{\rm g} = GM/c^2$ is the
gravitational radius. Inclination of the rotation axis of the black
hole to the line of site is given by $\mu_{\rm obs} \equiv \cos
\theta_{\rm obs}$.  We focus on black holes characterized by the
highest expected value of the spin parameter, $a=0.998$ (Thorne
1974). Such black holes are referred to in this paper as  maximally
rotating. Processes taking place in the ergosphere, which is a region
contained within the surface given by $R_{\rm erg} = R_{\rm g}[1+(1 -
a^2 \cos^2 \theta)^{1/2}]$ but outside the event horizon, appear
particularly important for our study.

We assume that the accreting material forms a geometrically thin,
optically thick disc and that a hot coronal region exists in the
neighborhood of the innermost part of the disc. We consider two models
of a disc-corona system:

\noindent
(i) an inner corona replacing a cold disc inside some transition
radius, $r_{\rm tr}$;

\noindent
(ii) a cold disc extending down to the black hole horizon with a hot
corona formed above the surface of the innermost disc; the radius of a
region covered by the corona in this model is denoted as $r_{\rm c}$.

We focus our investigation on the impact of strong gravity effects on
formation of spectra rather than on a precise modeling of a
disc-corona system. Therefore, we make a number of simplifying
assumptions in our model. 

We follow standard description for a geometrically thin disc
(e.g.\ Novikov \& Thorne 1973), assuming that the disc rotates in the
equatorial plane of the Kerr geometry with Keplerian angular velocity
(Bardeen et al.\ 1972),
\begin{equation}
\Omega_{\rm K}(r) = {1 \over a+r^{3/2}},
\end{equation}
and that it emits released energy as blackbody radiation with the
surface brightness distribution described by the standard formula
given by Page \& Thorne (1974). For a disc extending down to the black
hole [model (ii)] there is  no emission below the marginally stable
orbit, located at $r_{\rm ms}=1.23$ for $a=0.998$.

In our model of a corona  we assume that hot electrons have Maxwellian
distribution of energy  in the corona rest frame; that the corona is
isothermal; and that the coronal material corotates with the disc
(except for model B defined below) with no transversal or radial
motion, $u^{\theta} = 0$ and $u^r = 0$. 

While corotation of the hot plasma with the disc seems natural for a
model with corona formed above the disc surface, velocity field  in
the inner corona is an uncertain issue. However, hydrodynamical
studies typically  indicate that in the extreme Kerr metric  the
region where radial velocity becomes relativistic  is very narrow,  
which supports our approximation of the velocity
field by a quasi-Keplerian motion.

In particular, solutions of advection-dominated accretion flows
(Popham \& Gammie 1998; Manmoto 2000) indicate that for a rapidly
rotating black hole  accretion proceeds with a rather low radial
velocity  even close to the event horizon (while for slowly rotating
black holes rapid acceleration of the flow  occurs at further
distances).  Although these results are not directly applicable to our
model as they are based on assumption that a quasi-spherical flow is
formed far from a black hole, they confirm that fast rotation of a
black hole stabilizes  rotationally supported  flows.  Specifically,
for $a=0.998$ these models typically yield radial velocity increasing
from  $0.3c$ at $r=2$ to $0.7c$  at $r=1.2$.  Emission of accretion
flows with such velocities differs from emission of a Keplerian flow
mainly by increased fraction of photons captured by black hole (see
Section 3.2). Then, overestimated strength of emission  from the
ergospheric region is a major shortcoming due to neglect  of radial
velocity in modeling spectra of such flows.

Even stronger support  for  our model comes from the MHD simulations,
which do not reveal formation of any  significant sub-Keplerian
component. In simulations with $a = 0.998$ a quasi-Keplerian
torus extends deep into the ergosphere and the plunging region is
extremely narrow (the radial velocity profile on Figure 12 in De Villiers, 
Hawley \& Krolik 2003 shows velocity smaller than $0.1c$ at $r_{\rm
ms}$).

On the other hand, our description of the accretion flow is not appropriate
for investigation of spectra produced by strongly sub-Keplerian flows,
which are postulated in models considering Comptonization
dominated  by transfer of  kinetic energy from bulk motion of such a
flow  (e.g.  Chakrabarti \& Titarchuk 1995, Laurent \& Titarchuk
1999). We make a thorough study of spectra arising from such
relativistic inflows in a separate paper, extending formalism
developed here by a proper treatment of the propagation of photons in
a plasma with high radial velocity.

Velocity of the corona rest frame
relative to the LNR frame and corresponding Lorentz factor are 
related to angular velocity, $\Omega(r,\theta)$, by (Bardeen et al.\
1972)
\begin{equation}      
V \equiv {v^\phi \over c} = { \sin{\theta} A \over \Sigma
   \Delta^{1/2} } \left(  \Omega - \omega \right),~~~~~ \Gamma_{\rm
   ln} = \left[ 1 - V^2 \right]^{-1/2},
\label{gamma}
\end{equation}
where
\[
\Delta=r^2 - 2r +a^2,~~~\Sigma= r^2 + a^2 \cos^2 \theta,
\]
\begin{equation}
A=(r^2 + a^2)^2 - a^2 \Delta \sin^2 \theta,~~~\omega = 2 a r /A.
\label{delsig}
\end{equation}

The rest density of electrons in the corona, $n$, is described by  
a dimensionless parameter,
\begin{equation}
\hat n = n R_{\rm g} \sigma_{\rm T},
\label{nhat}
\end{equation}
where $\sigma_{\rm T}$ is the Thomson cross-section.

The following assumptions on distributions of the rest density and
angular velocity define three specific models of the corona.

\noindent
{\bf Model A:} a spherical  corona.  This model assumes that the
distribution  of rest density is uniform, $\hat n = {\rm const}$, in a
sphere with radius $r_{\rm tr}$ for model (i)  and $r_{\rm c}$ for
model (ii), and that the angular velocity is constant on spheres, with
\begin{equation}
\Omega(r,\theta) =  \Omega_{\rm K}(r).
\label{ang_vel}
\end{equation}

\noindent
{\bf Model B:} a non-rotating  spherical corona. This model adopts the
same assumptions as model A except for angular velocity which is set
to be equal to the velocity of the LNR observer, $\Omega = \omega$,
yielding $V \equiv 0$.

\noindent
{\bf Model C:} a slab-like corona. This model assumes that angular
velocity distribution is given by
\begin{equation}
\Omega(r,\theta) =  \left[ a+(r \sin \theta)^{3/2} \right]^{-1}
\label{omega2}
\end{equation} 
and that the corona fills only a region within $12 \degr$ from the
equatorial plane. The latter assumption is due to the fact that
angular velocity given by equation (\ref{omega2})  violates the
causality condition,  $g_{tt} + 2 g_{t \phi} \Omega + g_{\phi \phi}
\Omega^2< 0$, at high latitudes.  In this model we assume also that
the rest density decreases with radius as $\hat n \propto r^{-1}$,
which gives approximately constant vertical optical depth of the
corona.

Most of our model spectra, in both geometries of a disc-corona system, 
are computed for a spherical corona model A. Spectra of discs sandwiched 
by coronae described by models B and C are presented in Section 4.2 
to illustrate influence of assumed velocity field and density
distribution on spectral properties.

We do not attempt to model a region where the flow accelerates inward
to relativistic velocities. As discussed above, this region  should be
very narrow and its emission is very unlikely to give any  noticeable
contribution to observed spectrum. Specifically,  we neglect the
region inside $r_{\rm min}=1.13$, where approximation by a Keplerian
motion would yield artificially high orbital velocity ($V > 0.6$,
approaching the speed of light at the prograde circular photon orbit
located at $r_{\rm ph} = 1.07$). We have checked  that spectra
presented in this paper are weakly (in most cases negligibly)
sensitive on the choice of any specific value of $r_{\rm min}$ close
to $r_{\rm ms}$.  The most significant change of spectrum due to the
change of this parameter occurs for edge-on spectrum in Fig.\
\ref{fig9}(a), which has a strong contribution of emission from deep
regions of the ergosphere (spectra obtained with $r_{\rm min} = 1.13$
and 1.23 are compared in  that figure).

\subsection{Photons in the Kerr geometry}

We summarize here relevant properties of photon motion in the Kerr
metric. A photon trajectory is determined by two dimensionless
constants of motion
\begin{equation}
\eta \equiv {Q c^2 \over E_{\rm inf}^2 R_{\rm g}^2},~~~ \lambda \equiv
{L c \over E_{\rm inf} R_{\rm g}},
\end{equation}
where $E_{\rm inf}$ is the photon energy at infinity, $Q$ is the
Carter's constant, $L$ is the component of angular momentum parallel
to the black hole rotation axis. The following equations govern the
photon trajectory (e.g.\ Bardeen et al.\ 1972)
\[
{{\rm d} {\hat t} \over {\rm d} \zeta} =  {A(1 - \omega \lambda) \over
                  \Sigma \Delta},~~~ 
{{\rm d} r \over {\rm d} \zeta} = \pm {R^{1/2} \over \Sigma},~~~ 
{{\rm d} \theta \over {\rm d} \zeta} =  \pm {\Theta^{1/2}
                  \over \Sigma},
\]
\begin{equation}
{{\rm d} \phi   \over {\rm d} \zeta} = { \lambda (\Sigma - 2r)  \over
         \Sigma \Delta \sin^2 \theta} + {2ar \over \Sigma \Delta},
\label{trtp}
\end{equation}
where
\[
R = \left( r^2 +a^2 - \lambda a \right)^2 - 
        \Delta  \left[ (\lambda - a )^2 + \eta \right],
\]
\begin{equation}
\Theta = \eta + 
   \cos^2 \theta \left( a^2 - \lambda^2/\sin^2 \theta \right),
\end{equation}
and $\zeta$ is an affine parameter.

While photon motion is simply described in the BL coordinates, local
rest frames are more appropriate for description of physical
processes. Applying the transformation between the BL coordinate frame
and the LNR frame, given by Bardeen et al.\ (1972), we find energy and
momentum of a photon measured by the LNR observer
\[
E_{\rm ln} = E_{\rm inf} \left(1 - \omega \lambda  \right) \left( {A
   \over \Sigma \Delta} \right)^{1/2},~~~ p_{(\phi)}   = {E_{\rm inf}
   \over c}{\lambda \over \sin{\theta} }  \left( {\Sigma \over A}
   \right)^{1/2},
\]
\begin{equation}
p_{(\theta)} = \pm {E_{\rm inf} \over c} \left( {\Theta \over \Sigma}
    \right)^{1/2},~~~ p_{(r)}    = \pm  {E_{\rm inf} \over c} \left(
    {R \over \Sigma \Delta} \right)^{1/2}.
\label{energy}
\end{equation}
Then, a Lorentz transformation  yields the energy and momentum in the
rest frame of the plasma. In particular, photon energy in the rest
frame
\begin{equation} 
E_{\rm rest}  = \Gamma_{\rm ln}  \left( E_{\rm ln} - v^{\phi}
      p_{(\phi)} \right).
\label{erest}
\end{equation}

Putting in the above equations velocity of a Keplerian disc, $v^{\phi} = V_{\rm
K}  \equiv (\Omega_{\rm K} - \omega) A /( \Sigma \Delta^{1/2})$, we find
direct relations (given in a similar form by Bardeen \& Cunningham
1973) between the constants of motion of a photon, $\lambda$, $\eta$
and $E_{\rm inf}$, and its energy and directional angles measured in
the disc rest frame
\[ 
g \equiv {E_{\rm inf} \over E_{\rm disc}} =  \left( {\Delta \Sigma
\over A} \right)^{1/2} { \left( 1 - V_{\rm K}^2 \right)^{1/2} \over 1
-  \Omega_{\rm K} \lambda} ,~~~\eta=\left( {r \cos \theta_{\rm em} \over g}
\right)^2,
\]
\begin{equation} 
\lambda = {\sin \phi_{\rm em} \sin \theta_{\rm em} + V_{\rm K} \over
\Omega_{\rm K} \sin  \phi_{\rm em} \sin \theta_{\rm em} + \Sigma
\Delta^{1/2}A^{-1} \ + \omega V_{\rm K}},
\label{gle}
\end{equation}
where $E_{\rm disc}$ is the energy in the disc rest frame,
$\theta_{\rm em}$ is the polar angle with respect to direction
perpendicular to the disc surface, and $\phi_{\rm em}$ is the
azimuthal angle, in the disc plane, with respect to coordinate $r$
direction.

For motion in a curved space-time we must integrate optical depth
along a photon trajectory separately for each photon. The increase of
the optical depth is given by ${\rm d} \tau = \sigma(E_{\rm rest}) n
{\rm d} l'$, where $\sigma$ is the Klein-Nishina cross-section
(averaged over Maxwellian distribution of electron velocities), $n$ is
the rest density of electrons and ${\rm d} l'$ is the length along the
photon path measured in the plasma rest frame. We find this length
using the obvious relation, ${\rm d} l' = c {\rm d} t'$, where ${\rm
d} t'$ is the time interval measured in the plasma rest frame which
can be derived for a given trajectory as follows. Transforming the BL
coordinate differentials ${\rm d} \hat t$ and ${\rm d} \phi$ to the
LNR frame and then making a Lorentz transformation to the plasma
rest frame we get
\begin{eqnarray} 
{c  {\rm d} t' \over R_{\rm g}} & = & \Gamma_{\rm ln}  \left[(\Delta
\Sigma / A)^{1/2} {\rm d} \hat t  \right. \nonumber \\ & & \left. - V
(A / \Sigma)^{1/2} \sin{\theta}  \left( {\rm d} \phi - \omega {\rm d}
\hat t \right) \right].
\end{eqnarray}
Then, using equation (\ref{nhat}), we obtain
\begin{eqnarray} 
{\rm d} \tau & = &  \hat n {\sigma(E_{\rm rest}) \over \sigma_{\rm T}}
     \Gamma_{\rm ln} \left( {\Delta \Sigma \over A} \right)^{1/2}
     \left[ \left( 1 + V { 2 a r \sin{\theta} \over \Sigma
     \Delta^{1/2}} \right) {{\rm d} \hat t \over {\rm d} \zeta}
     \right.  \nonumber \\ 
 & & \left.   - V { A \sin{\theta}
     \over \Sigma  \Delta^{1/2}}{{\rm d} \phi \over {\rm
     d} \zeta} \right] {\rm d}\zeta,
\label{tau}
\end{eqnarray}
where the derivatives ${\rm d} \hat t / {\rm d} \zeta$ and  ${\rm d}
\phi / {\rm d} \zeta$ are given by equation (\ref{trtp}).

\begin{figure}
\includegraphics[height=90mm]{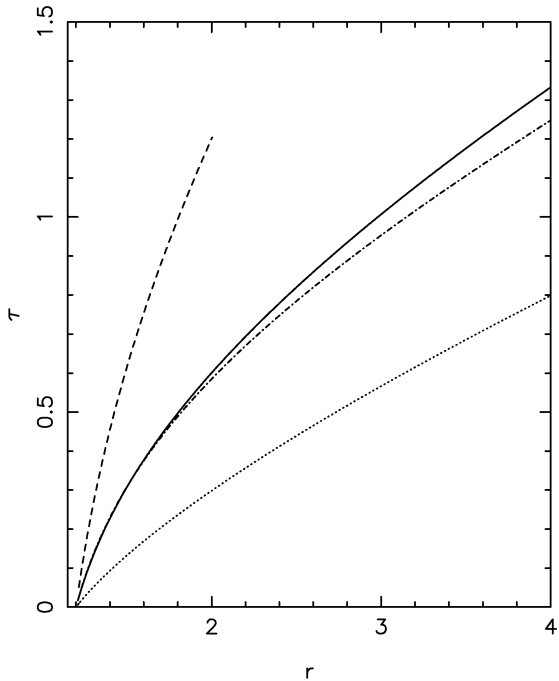}
\caption{Optical depths along radial trajectories of outgoing photons
emitted at $r=1.2$.  Solid, dotted and dashed curves correspond to
trajectories in the equatorial plane of a Kerr black hole ($a=0.998$).
Dot-dashed curve is for trajectory with a constant polar angle
$\theta=0.1\pi$ (close to the polar axis).  Photons propagate through
a thermal plasma with $kT_{\rm e}=80$ keV and with the rest density
$\hat n = 0.4$ for the dashed curve and $\hat n = 0.2$ for the
remaining curves.  Dotted curve is for a photon with $E_{\rm inf}=50$
keV, the remaining curves for $E_{\rm inf}=0.1$  keV. Note that the
optical depth for a photon with $E_{\rm inf}=50$ keV is reduced by 50
per cent between $r=1.2$ and  $r=2$ (compare dotted and solid
curves).}
\label{fig1}
\end{figure}

Note that in the ergosphere $E_{\rm rest}$ is typically a few times
higher than $E_{\rm inf}$ due to the gravitational blueshift.
Therefore, close to a black hole the Klein-Nishina corrections can
decrease significantly the optical depth  even for photons which in
flat space-time would be still in the Thomson regime. Fig.\ \ref{fig1}
shows optical depths integrated along radial trajectories
(characterized by $\lambda = 0$ and $p_{\theta}=0$; the latter
condition implies $\eta = -a^2 \cos^2 \theta$) of photons escaping
from $r=1.2$. Photons propagate  through a plasma with the same
temperature and rest density as assumed for models presented in
Section 4. Note that the optical depth for a photon with $E_{\rm
inf}=50$ keV is decreased by around 50 per cent in the ergosphere due
to the reduction of the cross section.  Then, curvature in spectra 
may occur at a few tens of keV (indeed seen
in some spectra presented in Section 4). 

Note also that assumptions underlying a spherical corona model yield
similar optical depths of the corona in different directions with
respect to the symmetry axis of  the system.

\subsection{Monte Carlo simulation}
Our procedure for computing spectra is based on the Monte Carlo method
and involves calculation of many photon trajectories. A similar
approach to computing spectra of black hole accretion flows was
applied, e.g., by Laor, Netzer \& Piran (1990), Laurent \& Titarchuk
(1999) and Kurpiewski \& Jaroszy\'nski (1999).  The free parameters of
our model are $r_{\rm tr}$ or $r_{\rm c}$ (depending on the
disc-corona model),  $a$, $M$, $\dot m$, $\hat n$ and electron
temperature, $T_{\rm e}$. Note that $M$ and $\dot m$ are relevant
only for the emission of the cold disc.  We trace individual photon
trajectories originating at the disc surface. The procedure for
generation of disc photons is described in Section 2.3.1 below.

Trajectories of photons reaching the hot plasma are followed, through
subsequent scatterings, according to algorithm described in Section
2.3.2. Photons reaching infinity are summed over 10 bins in $\mu_{\rm
obs}$ (from 0-0.1 to 0.9-1; $\mu_{\rm obs}$  given for spectra in
Section 4 correspond to the value in the middle of the bin). 
Photon transfer equations are solved down to $r_{\rm
min} = 1.13$.  Photons crossing the sphere of this radius 
are treated  as captured by the black hole.  

Spectra presented in Section 4 have been obtained in Monte Carlo runs
involving at least $10^8$ photons scattered in the corona.

\subsubsection{Thermal disc emission}
The emission point, $r_{\rm em}$, on the disc surface is generated
according to radial emissivity, $F^+$, given by Page and Thorne
(1974). Consecutive photons originating from this $r_{\rm em}$ are
generated until their summed energy exceeds some fixed value, $E_{\rm
tot}$, much higher than a characteristic photon energy. When $E_{\rm
tot}$ is exceeded we generate next $r_{\rm em}$.

For each photon its energy, $E_{\rm disc}$, is generated from the
blackbody distribution with the local surface temperature, $T_{\rm
surf} = (F^+/\sigma)^{1/4}$.  The initial direction in the disc rest
frame is generated with a uniform distribution in $\phi_{\rm em}$. The
polar angle, $\theta_{\rm em}$, is generated according to angular
distribution of local disc emission, with the specific intensity
$I(\cos \theta_{\rm em}) \sim 1 + 2.06 \theta_{\rm em}$ corresponding
to limb darkening in electron scattering limit (see, e.g.,
Gierli\'nski, Macio\l ek-Nied\'zwiecki \& Ebisawa 2001). Constants of
motion, $\eta$, $\lambda$ and $E_{\rm inf}$, are found from equation
(\ref{gle}). Then, we solve $r$ and $\theta$ equations of the set of
equations (\ref{trtp}) and determine whether the photon escapes to
infinity (contributing to the  spectrum of the disc emission), returns
to the disc (we neglected here  increase of  the disc temperature due
to these returning photons), illuminates the corona or gets captured
by the black hole.

\subsubsection{Comptonization}

If a photon reaches the coronal region, the optical depth is generated
according to the decreasing exponential distribution $P(\tau) = {\rm
e}^{-\tau}$. Then, we solve equations (\ref{trtp}) and integrate the
optical depth,  given by equation (\ref{tau}), until the trajectory
crosses the horizon (see Section 2.3.3), leaves the corona or reaches
the chosen optical distance. In the last case a simulation of Compton
scattering is performed as follows:

\noindent
(i)  Photon momentum and energy in the plasma rest frame are found, as
described in Section 2.2.

\noindent
(ii) Electron velocity is generated  and scattered photon direction
and energy are derived according to the procedure described in
G\'orecki \& Wilczewski (1984).

\noindent
(iii) The four-momentum of the scattered photon is Lorentz-transformed
back to the LNR frame, then equations (\ref{energy}) are inverted for
new constants of motion ($E_{\rm inf}$, $\eta$, $\lambda$).

Photons leaving the corona are treated  similarly to those emitted
from the disc surface - considering their motion  in the $r\theta$
plane we determine whether they escape from the system and contribute
to the Comptonization spectrum.

\section{Compton scattering in the Kerr metric}

\subsection{Single scattering spectra}

Photons scattered in a hot corona gain energy due to both thermal
motion of electrons and bulk motion of the corona. Furthermore,
dynamics of individual scatterings  is influenced by gravitational
energy shifts.  In this section we determine the range of distances
from a black hole in which the gravitational effects are important for
the formation of Comptonization spectra.  Then, we examine
qualitatively how these effects can affect the emerging spectra.  For
this investigation we derive spectra of a single scattering of
monoenergetic photons, with initial energy $E^0_{\rm inf}$, on
electrons in the Kerr geometry. We assume various distributions of
electron energies to compare effects taking part in formation of the
spectra.   Simulation of Compton scatterings is performed according to
procedure  described in Section 2.3.2.

\begin{figure}
\includegraphics[height=85mm]{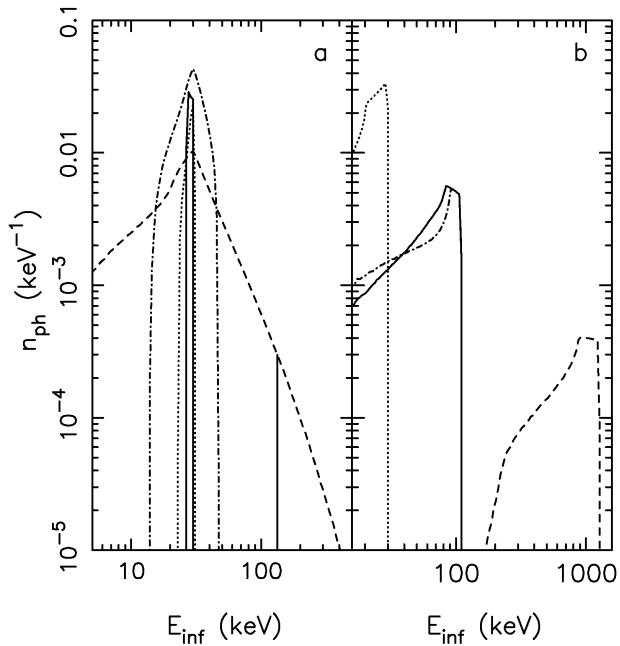}
\caption{Angle-averaged spectra for a single scattering of photons,
with $E^0_{\rm inf}=30$ keV, impinging on electrons at rest (with
respect to the LNR frame) in the equatorial  plane of the maximal Kerr
geometry.  {\bf (a)} Photons illuminate electrons isotropically (as
seen in the LNR frame).  Scatterings take place at $r=1.4$ (dashed
curve), $r=3$ (dot-dashed curve) and $r=6$ (dotted curve). Solid
curve gives the spectrum for a single scattering on electrons at rest
in flat space-time, integrated over all scattering angles. The solid
and dotted curves are rescaled by a factor of 0.1.  Photons emerging
from $r=1.4$ with energies exceeding 135 keV (indicated by the solid
vertical line) come from  scatterings which extract the black hole
rotational energy.  {\bf (b)} Dependence of the spectrum for
scattering at $r=1.4$  on the direction of impinging photons. The
initial direction of photons is given by the following angles (defined
in the text; $\mu_{\rm ln} \equiv \cos \theta_{\rm ln}$):  $\mu_{\rm
ln}=0$, $\phi_{\rm ln}=\pi$ for dot-dashed curve;  $\mu_{\rm ln}=0$,
$\phi_{\rm ln}=\pi/2$ for dotted curve; $\mu_{\rm ln}=1$ (which
uniquely determines the direction) for solid curve; and $\mu_{\rm
ln}=0.2$, $\phi_{\rm ln}=1.91\pi$ for dashed curve. Scatterings
extracting the rotational energy take place only for photons impinging
from the last direction.}
\label{fig2}
\end{figure}

Spectra presented in this section are normalized by the number of
impinging photons, $n_{\rm ph}(E_{\rm inf}) \equiv N(E_{\rm inf})/
N_0$, where $N_0$ is the number of incident photons and $N(E_{\rm
inf})$ is the number of photons observed, after scattering, at
infinity with energy $E_{\rm inf}$.  Note that photons captured by a
black hole  do not contribute to the spectrum, therefore
\[
\int n_{\rm ph}(E_{\rm inf}) {\rm d} E_{\rm inf} < 1.
\]

\begin{figure}
\includegraphics[height=68mm]{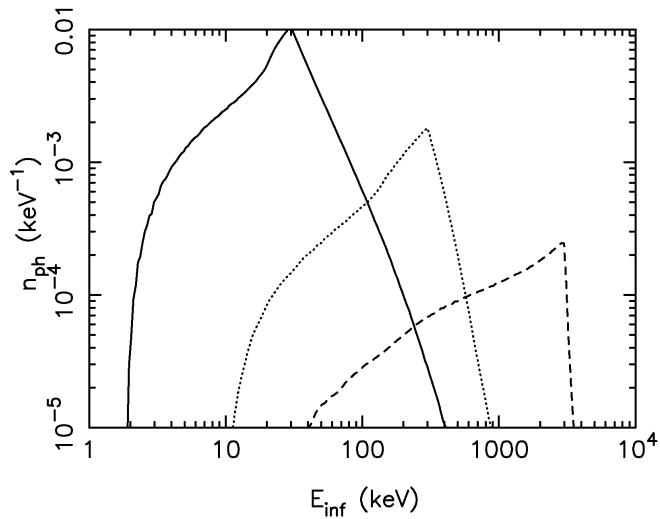}
\caption{Angle-averaged spectra for a single scattering, at $r=1.4$,
on electrons at rest in the equatorial plane of the maximal Kerr
geometry. Solid, dotted and dashed curve  is for  $E^0_{\rm inf}=30$,
300 and 3000 keV, respectively.  Photons illuminate electrons
isotropically, as seen in the LNR frame.}
\label{fig3}
\end{figure}

To illustrate effects of general relativity, separated from kinematic
and thermal effects, we examine first scattering on electrons at rest
with respect to the LNR frame.   Fig.\ \ref{fig2}(a) shows spectra for
scattering of  photons, with $E^0_{\rm inf}=30$ keV, on electrons
located  in the equatorial plane of a Kerr black hole.  We assume that
photons illuminate electrons isotropically  (as seen in the LNR frame)
excluding directions which correspond to negative energy, $E^0_{\rm
inf} < 0$, of impinging  photons.  Also shown is the spectrum,
integrated over all scattering angles, for scattering of photons with
energies $E_0=30$ keV on electrons at rest in  flat space-time.  For
$E_0 \ll m_{\rm e} c^2$ Compton scattering is nearly elastic and the
spectrum emerging in flat space-time is very narrow.  On the other
hand,  scattering in the Kerr space-time produces spectra extending
remarkably further into both low and high energies. The low energy
part of such spectra is a generic feature of scattering close to a
black hole,  resulting from the gravitational blueshift of incoming
photons. Due to this blueshift, close to a black hole photons lose in
a scattering, on average, a larger amount of energy than in flat
space-time.

More importantly,  in the Kerr metric the relation between photon
energies in the LNR frame and at infinity involves photon angular
momentum, see equation (\ref{energy}). Therefore, the  photon  energy
after scattering may be higher than its energy before scattering,
provided that a suitable change of angular momentum occurs.  Such
effect is not  allowed in either flat space-time or the Schwarzschild
metric.  The increase of photon energy can be particularly significant
in the black hole ergosphere, where Penrose Inverse Compton
scatterings (Piran \& Shaham 1977a), for which a scattered photon has
higher energy than the sum of the photon and electron energies before
scattering, are permitted. Photons produced in such events form the
high energy part of the spectrum,  as indicated on Fig.\ \ref{fig2}(a)
for the spectrum from $r=1.4$, which extends to energies exceeding the
initial energy by over an order of magnitude.

The necessary condition for a Penrose scattering to take place is a
very large  blueshift of an impinging photon in the LNR frame.  The
blueshift is determined by the angle, $\Phi_{\rm ln}$, between the
photon  momentum and the $\phi$ direction in the LNR frame. Namely,
using equation (\ref{energy}),  with $p_{(\phi)} = (E_{\rm ln}/c) \cos
\Phi_{\rm ln}$, we get
\begin{equation}
E_{\rm ln} = E_{\rm inf} (\Sigma A)^{1/2}  \left(
\Sigma \Delta^{1/2} + 2ar \sin \theta \cos \Phi_{\rm ln} \right)^{-1}.
\label{eln}
\end{equation}
Increasingly high blueshifts  correspond to narrower ranges of initial
directions.  E.g., for $E_{\rm inf}=30$ keV and $r=1.4$, $E_{\rm ln}$
exceeds 100 keV for $0.45 \pi < \Phi_{\rm ln} < 0.59 \pi$, while
$E_{\rm ln}>1000$ keV for  $0.58 \pi < \Phi_{\rm ln} < 0.59 \pi$.
Note also that strongly blueshifted  photons have very large negative
angular momentum ($\lambda \ll 0$) and can emerge only from a
radiative process taking place in the ergosphere.

Clearly, Penrose scatterings are permitted only for certain initial
directions of a photon.  Fig.\ \ref{fig2}(b) illustrates dependence of
the scattering spectrum on initial direction.  Note that while the
blueshift is determined by $\Phi_{\rm ln}$, different initial
directions with the same $\Phi_{\rm ln}$ give rise to different
scattering spectra at infinity.  The spectra in Fig.\ \ref{fig2}(b)
correspond to initial directions determined in the LNR frame by the
polar angle, $\theta_{\rm ln}$, between the photon initial direction
and coordinate $\theta$-direction, and the azimuthal angle, $\phi_{\rm
ln}$, in the $r \phi$ plane, with respect to the $r$-direction. These
angles are  related to $\Phi_{\rm ln}$ by   $\cos \Phi_{\rm ln} =
\sin \phi_{\rm ln} \sin \theta_{\rm ln}$.

\begin{figure}
\includegraphics[height=68mm]{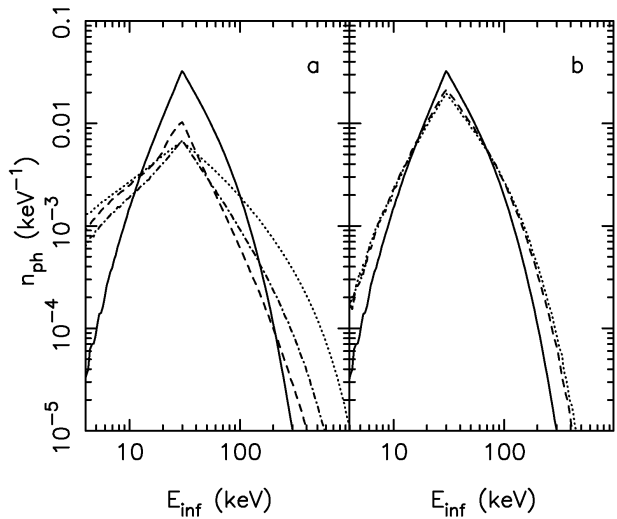}
\caption{A spectrum for a single scattering of photons, with $E_0=30$
keV, on electrons with a Maxwellian distribution of velocities,
$kT_{\rm e}=80$ keV, in flat space-time (given on both figures by the
solid curves) compared with angle-averaged spectra  for a single
scattering of photons, with $E^0_{\rm inf}=30$ keV, on electrons in
the equatorial plane  of the Kerr geometry. {\bf (a)} Scattering at
$r=1.4$.  Dashed curve is for scattering on cold electrons at rest in
the LNR frame.  Dot-dashed curve is for scattering on thermal
electrons, with $kT_{\rm e}=80$ keV, at rest  in the LNR frame.
Dotted curve is for scattering on thermal electrons ($kT_{\rm e}=80$
keV) on a Keplerian orbit.  {\bf (b)}
Scattering at $r=6$ for $a=0$ (dashed curve) and $a=0.998$ (dotted
curve).  Target electrons have thermal energies ($kT_{\rm e}=80$ keV)
and rotate on a Keplerian orbit. Note that difference between spectra
emerging in the Schwarzschild and extreme Kerr geometries is
negligible.}
\label{fig4}
\end{figure}

Considering possibility of generating high energy photons due to 
Penrose processes in black hole  accretion flows, we note that a
fine tuning of the initial direction is required for the most
efficient scatterings and therefore probability of such  events is
low. Moreover, even if a photon propagates with sufficiently high
blueshift in a local rest frame, probability of a scattering is
strongly reduced due to the decline of the Klein-Nishina cross-section.

Fig.\ \ref{fig3} shows dependence of the scattering spectra on initial
energy of photons.  The fractional loss of photon energy, in the LNR
frame, increases with increasing  $E_{\rm inf}^0$. Therefore, for
higher energy photons the gravitational effects  are relatively less
efficient in increasing photon energies.

\begin{figure}
\includegraphics[height=87mm]{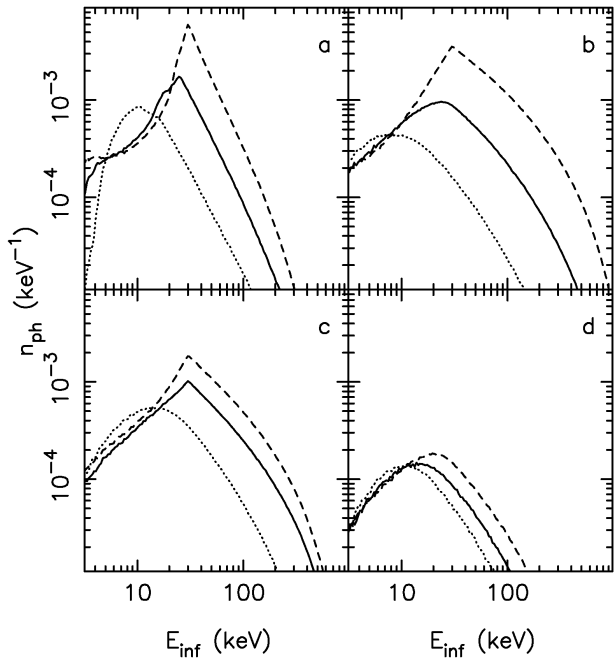}
\caption{Spectra for a single
scattering of monoenergetic photons, with $E_{\rm inf}^0=30$ keV,
illuminating isotropically (in the LNR frame)  electrons at $r=1.4$ in
the extreme Kerr metric.  Dashed, solid and dotted curves  give
spectra of photons scattered into $\mu_{\rm obs}=0-0.2$, $0.4-0.6$ and
$0.8-1$, respectively. {\bf (a)} Spectra for cold electrons at rest
in the equatorial plane; angle averaged spectrum is given by the
dashed curve on Fig.\ \ref{fig2}(a). {\bf (b)} Spectra for electrons, 
on a Keplerian equatorial orbit, with a Maxwellian distribution of 
velocities ($kT_{\rm e}=80$ keV) in the rest frame; angle averaged spectrum
is given by the dotted curve on Fig.\ \ref{fig4}(a). {\bf (c)} Spectra for
electrons with a Maxwellian distribution of velocities ($kT_{\rm
e}=80$ keV) in the rest frame. The rest of electrons rotates on a circular 
orbit, at $\theta=\pi/3$,
with angular velocity given  by equation (\ref{ang_vel}).  {\bf (d)}
Same as in Fig.\ \ref{fig5}(c) but the spectra are formed by photons
which do not cross the equatorial plane after scattering.}
\label{fig5}
\end{figure}

Outside the ergosphere, the strong gravity effects are not so
noticeable, as illustrated on Fig.\ \ref{fig2}(a).  Photons scattered
at $r=3$ achieve energies exceeding the initial energy by up to 50 per
cent and deviation of spectrum emerging from scatterings at $r=6$ from
the flat space-time scattering spectrum is insignificant. Therefore,
outside the ergosphere other effects should be dominant in formation
of Comptonization spectra.

Fig.\ \ref{fig4} compares efficiency of raising the energy  due to
gravitational shifts with boosting of photon energies resulting from
transfer of  thermal and bulk motion energies of electrons.  Note that
the probability that a photon with $E_{\rm inf}^0=30$ keV emerges with
energy exceeding 100 keV after scattering on a cold electron at
$r=1.4$  is equal to that corresponding to a single scattering in a
semi-relativistic thermal plasma in flat space-time.  Velocity of the
circular motion remains relatively low even in the deep ergosphere
($\Gamma_{\rm ln} \la 1.3$).  However, combined effects of strong
gravity,  thermal and bulk motions may result in a large gain of
energy in a single scattering, as illustrated by the dotted curve  on
Fig.\ \ref{fig4}(a).  Then, an optically thin plasma in the ergosphere
is able to generate photons with energies a few times higher than
thermal energy of electrons.

Outside the ergosphere, thermal  effects completely dominate.
Spectrum emerging from thermal electrons in circular orbit at $r=6$
differs only slightly from spectrum produced by the same thermal
population of electrons at rest in flat space-time [see Fig.\
\ref{fig4}(b)].  Moreover, the difference between the spectra
produced, at $r=6$, in the vicinity of a maximally rotating and a
non-rotating black hole is negligible.

Fig.\ \ref{fig5} illustrates dependence of spectra emerging from
Compton scattering  in the ergosphere on the inclination angle of a
distant observer.  Most photons emerging from the equatorial plane are
observed from directions close to that plane [see Fig.\ \ref{fig5}(a)]
due to  the bending of photon trajectories, which is the well known
effect of the Kerr geometry. Moreover, this gravitational focusing
affects primarily high energy photons. As a result,  spectra observed
edge-on are much harder than those observed face-on. Isotropic
distribution of  thermal velocities of electrons does not diminish
this anisotropy significantly,  on the other hand circular motion
further collimates radiation along the  equatorial plane [see Fig.\
\ref{fig5}(b)].

Slightly more isotropic spectra come from Compton scattering taking
place off the equatorial  plane [see Fig.\ \ref{fig5}(c)]. Moreover,
trajectories of  a significant part of photons  cross the equatorial
plane, as illustrated by Fig.\ \ref{fig5}(d).  In models with an
untruncated disc  such photons  are depleted from emerging spectra  as
they interact with the disc matter.

We find only a very weak anisotropy of spectra emerging from $r=6$,
regardless of the speed of black hole rotation.  Then, we conclude
that spectra of Comptonization in the Schwarzschild metric as well as
those  emerging at distances exceeding a few $R_{\rm g}$  in the Kerr
metric should be correctly approximated by flat space-time
Comptonization codes.

\begin{figure}
\centerline{\includegraphics[height=70mm]{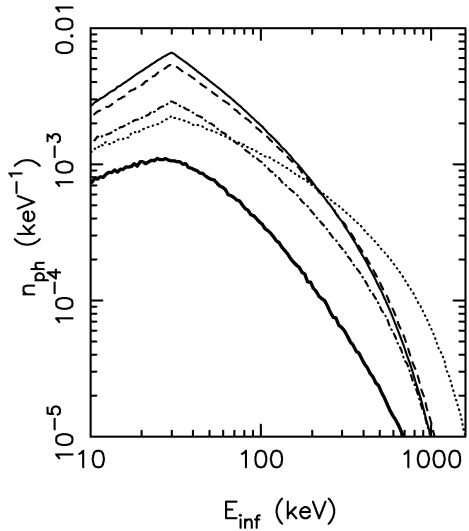}}
\caption{Angle-averaged spectra for a single scattering of
monoenergetic photons, with $E_{\rm inf}^0=30$ keV, on electrons 
with various velocities: $(v^r,v^{\phi})=(0,0.56c)$ for
thinner solid curve [Keplerian flow, same as  dotted curve on Fig.\
4(a)]; $(-0.3c,0.5c)$ for dashed curve; $(-0.7c,0.36c)$ for dot-dashed
curve; $(-0.9c,0)$ for thicker solid curve;  and $(-0.7c,0.56c)$ for
dotted curve.   Scatterings take place at $r=1.4$ in  the equatorial
plane. Isotropic illumination (in the LNR frame) of electrons and
Maxwellian distribution ($kT_{\rm e}=80$ keV) of electron velocities
in the rest frame are assumed.}
\label{fig6}
\end{figure}

\subsection{Radial velocity}

Fig.\ \ref{fig6} shows spectra for a single scattering, obtained
similarly  to spectra shown in Section 3.1, but for various velocity
fields of Comptonizing electrons. The velocity field is parametrized
by two components of velocity measured in  the LNR frame: the
orbital velocity, $v^{\phi}$, and the radial velocity,  $v^r$ ($=
\Delta^{-1} A^{1/2} u^r / u^t$; Bardeen et al. 1972).

The dashed and dot-dashed curves show spectra for values of $v^{\phi}$
and $v^r$  typical for solutions of advection-dominated accretion in
the ergosphere (see, e.g., Figure 1 in Popham \& Gammie 1998).
Clearly, scattering in a Keplerian flow (the thinner solid curve) results
in the highest fraction of escaping photons. On the other hand, highest
energies achieved by  scattered photons are similar in both models.

We emphasize that relatively high (Keplerian) orbital velocity is
essential for  spectra emerging from the innermost region of accretion
flow, as it collimates emission into directions for which escape to a
distant observer is possible.  The thicker solid and dot-dashed curves
illustrate how decrease of orbital velocity affects the
spectrum. Although the decrease of $v^{\phi}$ is connected with increase
of  $v^r$, yielding more efficient Doppler boosting ($\Gamma_{\rm ln} =
1.6$ and 2.3 for the dot-dashed and solid curve, respectively), it reduces 
efficiency of generating high energy photons able to escape.

Spectrum for the only velocity field producing photons more energetic
than a Keplerian flow (due to both a high Lorentz factor,
$\Gamma_{\rm ln} = 2.3$, and a Keplerian orbital velocity) is shown by
the dotted curve.  We emphasize, however, that such a velocity field is
very unlikely to describe a real flow as  total specific  energy in
such a flow would exceed by 30 per cent its rest energy, while the
opposite is expected in a dissipative flow.

We conclude that relativistic radial velocities, which can
significantly increase the Doppler boosting, are allowed
only in sub-Keplerian accretion flows, emission of which, however, is
beamed toward the event horizon and lost.

\subsection{High energy cut-off}

In flat space-time, repeated scatterings of soft photons in a
mildly-relativistic  thermal plasma lead to formation of a power-law
spectrum extending up to $kT_{\rm e}$.  Photons scattered to higher energies
achieve Wien equilibrium with electrons and the spectrum has a roughly
exponential turnover  reflecting the thermal  distribution of electron
velocities.  Close to a black hole this property is disturbed by
energy shifts between consecutive scatterings.  In particular, for
photons blueshifted to energies significantly  exceeding $kT_{\rm e}$,
thermal motions of electrons become less significant and the spectrum
is formed primarily due to  the recoil effect.    This effect distorts
the high energy cut-off from the shape  characteristic for thermal
spectra, as we illustrate in this section.

Due to a  non-local nature of Comptonization, analysis of formation of
Comptonization spectra in a curved space-time is a very complex issue.
Below we apply a simplified treatment  to derive spectra of
consecutive scatterings of soft seed photons  off thermal electrons, with
$kT_{\rm e}=80$ keV, which are in circular motion corresponding to
model A.  We assume that all scatterings take place in a fixed point
($r=1.4$ and $\theta=0.45 \pi$) and that in each scattering photons
impinge from the same direction, given by  $\phi_{\rm ln} = \pi$ and
$\mu_{\rm ln} = 0$.  For each photon its initial energy, $E_{\rm
inf}$, is generated from a blackbody  distribution with $kT_{\rm bb} =
100$ eV. Then, the following steps are repeated  to simulate four
subsequent scatterings of this photon:

\noindent
(i) rest energy is found [equations (\ref{energy}) and (\ref{erest})] -
the assumed initial direction yields $E_{\rm rest} = 6 E_{\rm inf}$; 

\noindent
(ii) simulation of Compton scattering in the rest frame is performed;

\noindent
(iii) energy of scattered photon, $E_{\rm inf}$, is found and equations 
of motion are solved; photons escaping to infinity contribute to spectrum 
of given scattering order.

\begin{figure}
\centerline{\includegraphics[height=50mm]{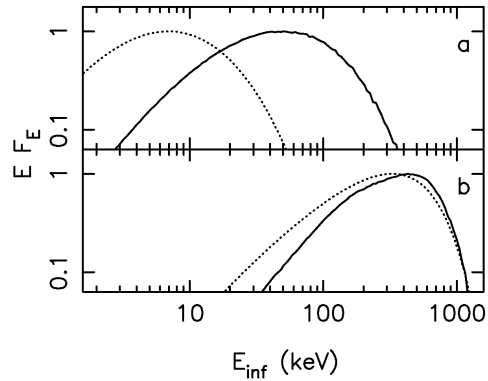}}
\caption{Solid curves show spectra emerging from second (a) and fourth
(b) scattering order, observed by a distant observer at $\mu_{\rm obs}
= 0.55$.  Scatterings take place at $r=1.4$ and $\theta=0.45 \pi$ (see
text for details).  Dotted curves show spectra, as a function of
photon energy at infinity,  before scattering.  All the spectra are
normalized to unity in maximum.}
\label{fig7}
\end{figure}

\begin{figure}
\centerline{\includegraphics[height=50mm]{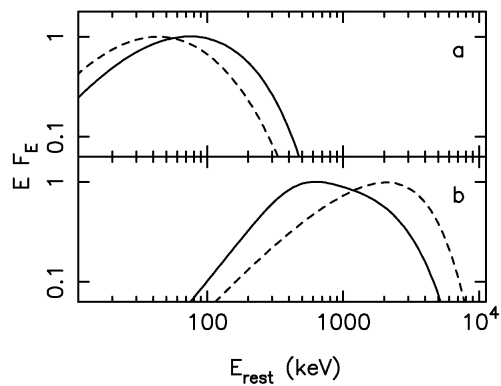}}
\caption{Figure illustrates increasing role of the recoil effect  in
formation of spectra emerging from higher scattering orders.  Solid
curves show angle-averaged spectra, in the rest frame, of photons
emerging in second (a) and fourth (b) scattering order.  Dashed curves
show spectra of photons before scattering.  }
\label{fig8}
\end{figure}

\noindent
Figs.\ \ref{fig7} and \ref{fig8} show spectra at infinity and in the 
rest frame, respectively, for second and fourth scattering orders.

The above procedure roughly approximates effects of photon transfer in
the Kerr metric.  Step (i) accounts for significant blueshifts
experienced by photons in the immediate vicinity of the event
horizon. In turn, the assumed initial direction  accounts for the fact
that bending  of photon trajectories results in very anisotropic
radiation field close to a black hole.  In simulations including full
treatment of photon propagation, presented in Section 4,  we find that
initial directions of photons scattered close to the event horizon are
concentrated along the inward radial direction (given by the assumed 
values of $\phi_{\rm ln}$ and $\mu_{\rm ln}$).

This anisotropic illumination is not important in formation of spectra
emerging from  lower scattering orders, for which the change of photon
energies in the rest frame is dominated by energy gain from Doppler
effect.  In such case, photons scattered in various directions have
similar spectra and the relativistic effects result in only a slight
redshift and smearing of the observed spectrum.  On the other hand,
for higher scattering orders the angular distribution of illuminating
photons becomes important, as spectra formed due to the recoil effect
depend on the  scattering angle and then photons scattered into
various directions are subjected  to different transfer effects, as
described below.

Due to high initial blueshift and moderate redshifts of photons
escaping after scattering  (typical redshift of escaping photons is
$0.5<g<1$), in lower scattering orders photon energies, $E_{\rm inf}$,
amplify by a large factor with each  scattering [see Fig.\
\ref{fig7}(a)], even though in the rest frame the fractional energy
transfer is small [Fig.\ \ref{fig8}(a)].

In higher scattering orders photons lose energy, in the rest frame,
forming spectrum shown by the solid curve on Fig.\ \ref{fig8}(b).
Spectrum of photons observed by a distant observer [solid curve on
Fig.\ \ref{fig7}(b)] is also affected by effects of photon transfer in
the Kerr space-time. In particular, scattered photons with $E_{\rm
rest} > 1$ MeV emerge from  forward scatterings and they weakly
contribute to the observed spectrum as most of them are captured by
the black hole.  Photons observed with $E_{\rm inf} <  200$ keV come
from backward scatterings and they are additionally redshifted by $g
\la 0.5$.

The recoil and transfer effects give rise to spectrum distorted from
the typical thermal shape, specifically, a flattened part occurs
between 200 and 400 keV in the spectrum on Fig.\ \ref{fig7}(b).
Contribution of higher scattering orders to Comptonization spectra
should result in similar distortion of the high energy cut-off,
although decreased scattering probability  resulting from the
(necessarily high) blueshift diminishes contribution of photons
scattered in this regime.  Spectral  breaks due to this effect are
indeed seen in spectra on Fig.\ \ref{fig9}(a) above 100 keV.

\section{Thermal Comptonization spectra}

Emergent spectra are shown in Figs.\ \ref{fig9}, \ref{fig10} and
\ref{fig12}-\ref{fig15}.  Each figure presents spectra for three
inclinations, $\mu_{\rm obs}=0.85$,  0.55 and 0.15.  The spectra have
been obtained for  stellar mass, $M=10 M_{\sun}$,  and supermassive,
$M=10^7 M_{\sun}$, black holes.  For a stellar black hole we assume a
distance of $d=5$ kpc  and for a supermassive one $d=5$ Mpc.  All
models assume a maximally rotating  black hole and $\dot m=0.5$.
Unless specified otherwise, we assume that corona is spherical (model
A) and that the electron temperature $kT_{\rm e}=80$ keV.  Motivated
by results of the previous section we focus on models with coronae
located within a few gravitational radii.

\subsection{Inner corona}
\begin{figure}
\includegraphics[height=85mm]{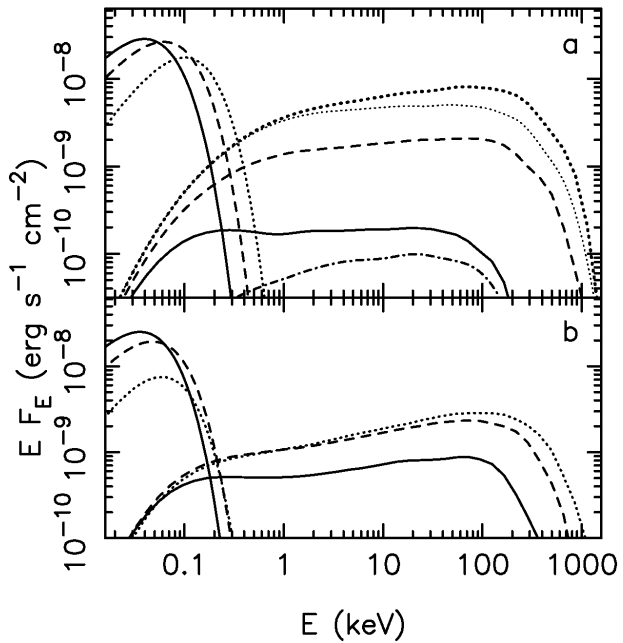}
\caption{Spectra emerging from a hot inner corona surrounded by a cold
accretion disc.  The model assumes a maximally rotating black hole
with a mass  $M=10^7M_{\sun}$, at a distance of $d=5$ Mpc, accreting
at $\dot m=0.5$. A uniform, spherical (model A, see Section 2.1) and
isothermal ($kT_{\rm e}=80$ keV) corona is assumed to replace the disc
within $r_{\rm tr}=2$ (a) and 4 (b). The rest density $\hat n=0.4$ and
0.2 for $r_{\rm tr}=2$ and 4, respectively.  The spectrum of thermal
emission from the cold disc and the Comptonized component are shown
separately. Solid, dashed and dotted curves correspond to  $\mu_{\rm
obs}=0.85$, 0.55 and 0.15, respectively. The dot-dashed curve in Fig.\
9(a) shows contribution, to spectrum observed at $\mu_{\rm
obs}=0.85$, of photons which experienced scattering inside $r=1.4$.
The thinner dotted curve in Fig.\ 9(a) shows spectrum for $r_{\rm min}=1.23$
(see Section 2.1).}
\label{fig9}
\end{figure}

\begin{figure}
\includegraphics[height=55mm]{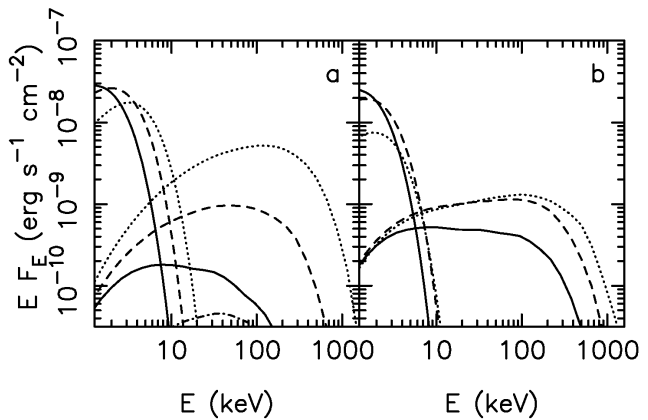}
\caption{Same as in Fig.\ \ref{fig9} but with $M=10M_{\sun}$ and $d=5$ kpc.
}
\label{fig10}
\end{figure}

Figs.\ \ref{fig9} and \ref{fig10} show spectra of models with a hot
inner corona replacing the disc within $r_{\rm tr}=2$ and 4.  We
assume values of  the rest density yielding  a similar Thomson optical
depth of the corona in both cases (see Fig.\ \ref{fig1}), namely $\hat
n=0.4$ for $r_{\rm tr}=2$ and $\hat n=0.2$ for $r_{\rm tr}=4$.  Two
main spectral components, emitted by the accretion flow, are shown
separately. Thermal emission of the outer disc, for the assumed  $\dot
m$, has a peak  around 0.1 keV and at a few keV for $M=10^7 M_{\sun}$
and $10 M_{\sun}$, respectively.  Comptonization  spectra emerging in
models with $r_{\rm tr}=2$ and 4 differ significantly, although the
hot plasma parameters are similar.

For $r_{\rm tr}=2$ virtually bulk of Comptonized radiation comes from
the ergosphere.  The luminosity, slope and cut-off energy of the
spectrum strongly depend on the inclination of the line of sight.
This is an obvious consequence of bending to the equatorial plane,
affecting trajectories of photons emerging from the ergosphere (see Fig.\
\ref{fig5}).

For photons observed close to the disc plane amplification of energy
in the first scattering is so large that their energies significantly 
exceed the energy of disc
photons.  As a result, a low energy cut-off of a Comptonization
spectrum may be clearly visible in the total spectrum [e.g., note
deficit of photons below 1 keV in the spectrum for  $\mu_{\rm
obs}=0.15$  on Fig.\ \ref{fig9}(a)].

The supermassive and stellar black hole models produce slightly
different spectra in the hard X-ray range.  This property is not
obvious as  in our model a change of $M$ affects directly only the
temperature  of the outer disk and hence the seed photons energy.  The
dependence of hard X-ray spectra on black hole mass is due to the fact
that photon transfer in the corona is affected by  gravitational
energy shifts (see Section 2.2).  As a result, the corona is optically
thin ($\tau \la 0.5$) for photons with energies of a few tens of keV,
emerging from the first scattering order in the stellar black hole
model,  but it is moderately thick ($\tau \ga 1$) for photons below 10
keV, which arise from first few orders  of scattering in the
supermassive black hole model.  Therefore,  higher energy density of
the radiation field establishes in the innermost region (where photons
achieve highest energies due to gravitational effects) of the corona
surrounding a supermassive black hole. Contribution  of high energy
photons generated in this region  results in the shape of the high
energy cut-off corresponding to the predominant role of the recoil
effect (Section 3.3) in the model with $M=10^7 M_{\sun}$.   Also,
stronger contribution of photons upscattered in the innermost region
[shown by dot-dashed curves on Figs.\ \ref{fig9}(a) and
\ref{fig10}(a)], but then scattered in the outer region  of the
ergosphere into directions corresponding to low inclination, results
in harder spectra (above a few tens of keV)  observed at $\mu_{\rm
obs} = 0.85$  in the model with $M=10^7 M_{\sun}$.

\begin{figure}
\includegraphics[height=90mm]{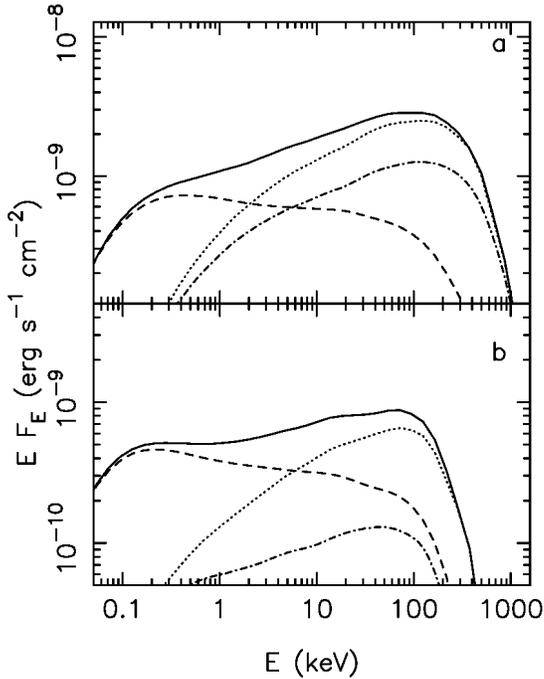}
\caption{ Solid curves show the Comptonization spectra of a model with
$M=10^7M_{\sun}$ and $r_{\rm tr}=4$ [same as in Fig.\ \ref{fig9}(b)]
for $\mu_{\rm obs}=0.15$ (a) and 0.85 (b).  Dotted and dashed curves
show contributions  of photons which experienced at least one
scattering in the ergosphere and photons scattered only outside the
ergosphere, respectively.  Dot-dashed curves show  contribution of
photons for which the last scattering took place in the ergosphere.}
\label{fig11}
\end{figure}

For $r_{\rm tr}=4$ most scatterings take place in a region where
standard thermal Comptonization effects dominate. However, the slope
and luminosity of the Comptonization component remain strongly
dependent on inclination.  This anisotropy is entirely caused by
different magnitude of the contribution of radiation produced in the
ergosphere to spectra observed from various directions (see Fig.\
\ref{fig11}).  Photons which do not  undergo scatterings in the
ergosphere  form spectra with the  photon spectral index
$\Gamma(1-80~{\rm keV}) \approx 2.1$.  A spectrum with such $\Gamma$
would result  from thermal Comptonization, with $kT_{\rm e} = 80$ keV and
$\tau=0.8 (= \hat n r_{\rm tr})$, in flat space-time.  This supports
our conclusion that the only remarkable influence of the space-time
metric  on the Comptonization spectrum occurs for radiation emerging
from the ergosphere.

Note also that ergospheric emission is significantly isotropized due
to  scattering in the outer regions of the corona.  Therefore,
contribution of radiation from the ergosphere to spectrum observed at
$\mu_{\rm obs} = 0.85$ is significant in model with $r_{\rm tr}=4$,
although photons escaping directly from the ergosphere contribute
negligibly to this spectrum.

As noted in Section 2.2, a spectral  curvature may occur at a few tens
of keV due to the decline of the Klein-Nishina cross-section.
Indeed, a spectral break between 10 and 20 keV, related to this
effect, is seen on Fig.\ \ref{fig11}(b).  Furthermore, emission from
the ergosphere has a significantly harder spectrum than emission from
the outer regions and their superposition may result  in an upturn of
the spectrum, such as seen on Fig.\ \ref{fig11}(b) around 1 keV.  Then,
a power-law approximation of the Comptonization spectrum may be not
adequate  for some regions of the parameter space.

Our model does not involve an explicit description of the corona
heating mechanism.  However,  the energy transferred to photons,
measured in the corona rest frame, determines  spatial distribution of
heating needed to maintain the assumed uniform distribution of
temperature. In particular, models with $r_{\rm tr}=4$ require that
around  40 per cent of the heating power should be provided to the
plasma within  the ergosphere.  Roughly 30 per cent of this energy  is
lost due to capturing of photons  by the black hole.  However, photons
upscattered within the ergosphere give a major contribution to the
total luminosity of Comptonized radiation, ranging from  50 per cent
at low inclinations to over 70 per cent in radiation observed edge-on.

\begin{figure}
\includegraphics[height=50mm]{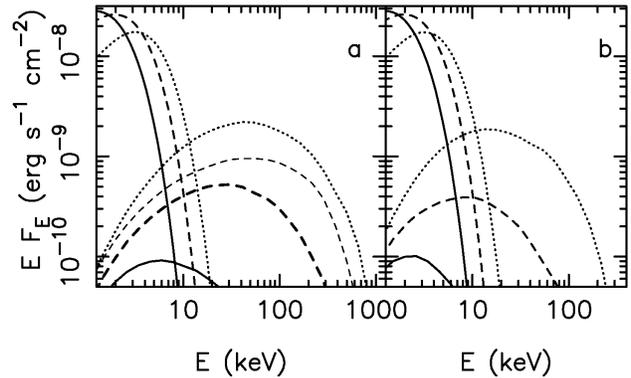}
\caption{Dependence of Comptonization spectra on the hot plasma
parameters  in a model with $r_{\rm tr}=2$ and $M=10 M_{\sun}$.   The
model parameters, except for $\hat n$ and $kT_{\rm e}$,  have the same
values as in Fig.\ \ref{fig10}(a).  {\bf (a)} $\hat n = 0.2$, $kT_{\rm
e}=80$ keV; for a comparison, Comptonization spectrum with $\hat n =
0.4$ [same as in Fig.\ \ref{fig10}(a)], for $\mu_{\rm obs}=0.55$,  is
shown by the thinner dashed curve  {\bf (b)} $\hat n = 0.4$, $kT_{\rm
e}=10$ keV.}
\label{fig12}
\end{figure}

Hardening of the Comptonization component, with increasing
inclination, is accompanied  by weakening of the disc
component. Thermal emission from the innermost region of the disc,
within several $R_{\rm g}$, is gravitationally focused to directions
close to the equatorial plane (Cunningham 1975; due to this effect,
emission from the disc truncated at $r_{\rm tr}=2$ is much stronger,
at high inclinations,  than emission from the disc with $r_{\rm
tr}=4$, see Figs.\ \ref{fig9} and \ref{fig10}). However, angular
distribution of emission from more distant regions of the disc is
mostly due to the reduction of the projected area,  with the observed
flux  given by $F_{\rm obs} \propto  \mu_{\rm obs}$ at $r \gg
1$. Then, the ratio of the Comptonization component to the disc
emission luminosities changes strongly with inclination.  E.g., for
$r_{\rm tr}=2$ the Comptonization component contributes 65 per cent to
the total luminosity at $\mu_{\rm obs}=0.15$ but only 5 per cent at
$\mu_{\rm obs}=0.85$.

Fig.\
\ref{fig12} illustrates dependence of the spectrum of ergospheric
emission on  electron temperature  and rest density in the
corona. Remarkably,  a relatively cold plasma ($kT_{\rm e}=10$ keV) is
able to produce spectra extending beyond 100 keV,  especially at high
inclinations [see Fig. \ref{fig12}(b)].

\begin{figure}
\centerline{\includegraphics[height=75mm]{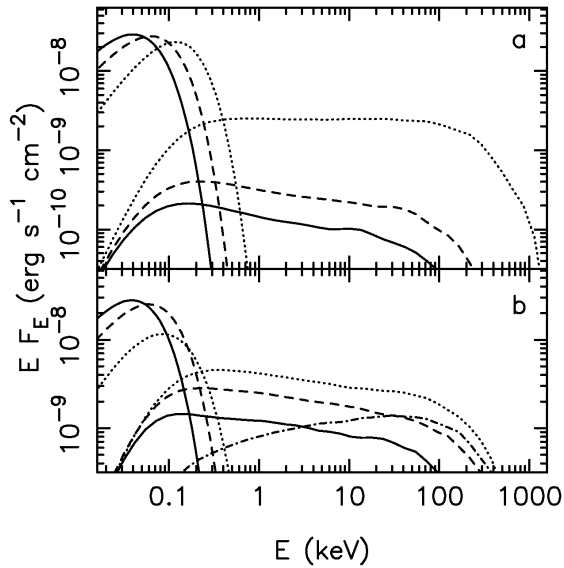}}
\caption{Spectra emerging from a spherical corona (model A)
surrounding the innermost disk surface with {\bf (a)} $r_{\rm c}=2$
and $\hat n=0.4$; {\bf (b)} $r_{\rm c}=4$ and $\hat n=0.2$. Other
parameters have values indicated in Fig.\ \ref{fig9}.  The dot-dashed
curve on Fig.\ \ref{fig13}(b) shows contribution of photons scattered
in the ergosphere to spectrum observed at $\mu_{\rm obs}=0.15$.}
\label{fig13}
\end{figure}

\begin{figure}
\includegraphics[height=50mm]{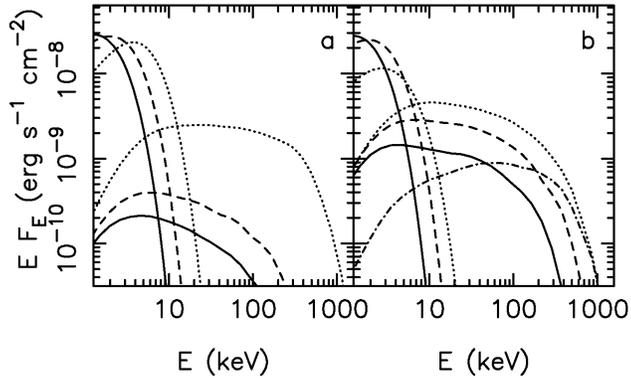}
\caption{Same as in Fig.\ \ref{fig13} but with $M=10M_{\sun}$ and $d=5$ kpc.
}
\label{fig14}
\end{figure}

\subsection{Corona above the disc surface}

Figs.\ \ref{fig13} and \ref{fig14} show spectra of models involving
an optically thick disc extending down to the event horizon, with  a
spherical corona surrounding the disc surface within $r_{\rm c}=2$ and
4.  The parameters of the corona are similar to these assumed above
for  the inner corona models, namely $\hat n=0.4$ for $r_{\rm c}=2$
and $\hat n=0.2$ for $r_{\rm c}=4$; $kT_{\rm e}=80$ keV. The innermost
disc immersed in the corona provides a very strong supply of seed
photons. However, the Comptonization spectra are softer than spectra emerging
from inner coronae with the same parameters, which is due to the fact
that    majority of photons upscattered in the ergosphere is absorbed
by the disc  [see Fig.\ \ref{fig5}(c)(d)].

In models with $r_{\rm c}=4$ the Comptonization spectra are dominated
by emission from regions outside the ergosphere and similar slopes
characterize spectra observed from various directions.  The
ergospheric component [shown on Figs.\ \ref{fig13}(b) and
\ref{fig14}(b)] contributes up to 25 per cent to the luminosity of
Comptonization radiation.  However, luminosity and cut-off energies of
the spectra increase for higher inclinations, which is due to the
circular motion of the corona.

\begin{figure}
\includegraphics[height=51mm]{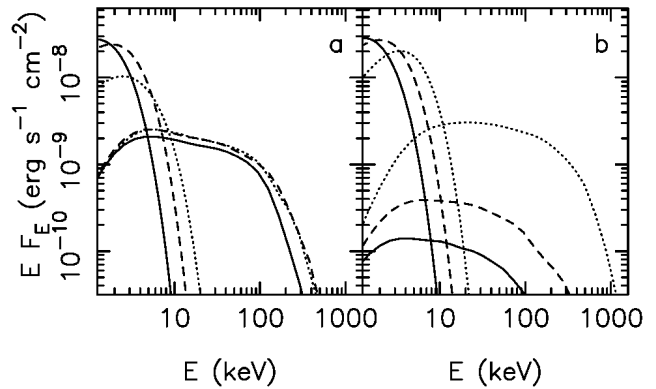}
\caption{ {\bf (a)} Same as in Fig.\ \ref{fig14}(b) ($r_{\rm c}=4$)
but for a nonrotating  spherical corona (model B).
{\bf (b)} Similar as in Fig.\ \ref{fig14}(a) ($r_{\rm c}=2$) but for a
corona described by model C with $\hat n = 4/r$.  }
\label{fig15}
\end{figure}

To confirm this last conclusion we show on Fig.\ \ref{fig15}(a)
spectrum of a non-rotating corona (model B).  The Comptonization
spectra are again dominated by emission generated outside the
ergosphere, which is weakly affected by the ray bending or
gravitational shifts of energies.  As a result, anisotropy of the
Comptonization component is negligible.

Finally, we consider sensitiveness of the properties of ergospheric
Comptonization spectra on angular velocity distribution in a corona.
Fig.\ \ref{fig15}(b) shows spectra emerging from a corona described by
model C with $\hat n = 4/r$. The corona fills a region within $12
\degr$ of the disc plane and covers the disc surface within $r_{\rm c}
= 2$.  In this model all scatterings take place close to the
equatorial plane in a plasma rotating with $V \approx 0.6$, which
results in slightly different angular distribution of Comptonized
photons with    respect to the spherical corona model. In the latter
model a large number scatterings take place in a slowly rotating
plasma at high latitudes.  However, the basic properties of the
spectra appear not affected by the change of velocity field and
geometrical shape  of the corona.

\section{Discussion}
\subsection{Comparison with similar studies}
The first study of the impact of the Kerr metric on spectra generated
by a black hole accretion flow was performed by Cunningham (1975). He
pointed out that most of the thermal emission from innermost regions
of a disc lying in the equatorial plane is  gravitationally  focused
into directions close to the disc plane.  We find that a similar
focusing affects also radiation arising from Comptonization in the
vicinity of a Kerr black hole.

Piran \& Shaham (1977a,b) investigated Compton scattering in the Kerr
metric.  They pointed out Compton scattering, taking place in the
ergosphere, as the most astrophysically feasible Penrose process.  As
an application of such processes they considered production of
$\gamma$-rays by a hot plasma in the ergosphere.

As we discuss in Section 3,  the gravitational effects may indeed
significantly increase the energy of a photon scattered in the
ergosphere.  However, probability of the most efficient processes  is
extremely low and they appear not important for formation of spectra.
Piran \& Shaham (1977b) point out dependence of the emerging spectra
on observer's inclination, in agreement with our results. On the other
hand, they concluded that thermal Comptonization in the ergosphere
gives rise to spectra  with a power-law shape extending into MeV range
without any  exponential-like cut-off.  We do not confirm this last
conclusion as our spectra, derived for similar electron temperatures,
have generic thermal-like cut-offs.  We suspect that this discrepancy
results from a rather poor statistical quality of spectra
presented by Piran \& Shaham (1977b).  Their spectra were obtained
with a typical number of $10^3$ photons.  On the other hand, we find
that at least $10^6$ photons scattered in a corona is needed in a
simulation to correctly derive the high energy part of angle-averaged 
spectrum.

Thermal Comptonization in two-temperature advection-dominated flows in
the Kerr metric was studied by Kurpiewski \& Jaroszy\'nski (1999) and
Manmoto (2000).  However, their models assumed a hot flow extending
out to $\sim 1000R_g$.  Therefore, their spectra arise mostly due to
thermal processes taking place far from a black hole, where
gravitational field is not important for radiative processes. Then,
these authors did not notice any direct impact of the gravitational
field metric on a Compton effect. A weak dependence of their spectra
on the black hole spin parameter is due to  changes in velocity field,
density and electron temperature profiles of the accretion flow.

\subsection{Applicability to black hole systems}

Clearly, effects considered in this paper are important  for the
emerging spectra only if a substantial part of hard X-rays comes from
the ergosphere.  On theoretical ground,  scenario with strong emission
from the ergosphere,  in case of accretion onto a rapidly rotating
black hole,  is supported by preliminary results of general
relativistic MHD simulations (see below)  as well as by the standard
theory of Keplerian $\alpha$-discs.  The latter predicts that   17 per
cent of the gravitational power is dissipated  within the ergosphere
(for $a=0.998$). This implies that significant emission of hard X-rays
from the ergosphere can be expected in models which assume that strong
release of the accretion energy occurs in a corona surrounding the
disc. Although details of the heat deposition (presumably involving
buoyancy and reconnection of magnetic fields) are uncertain, the
spatial  distribution of the heating is most likely to follow the
dissipation distribution in the disc.

Simulations of accretion driven by turbulent stresses generated by the
magnetorotational instability  (e.g.\ Hawley \& Balbus 2002; De
Villiers et al.\ 2003) support a model with an inner corona formed
within a few gravitational radii.  These simulations commonly indicate
formation of a hot inner torus, of the size slightly exceeding $r_{\rm
ms}$,  at the inner edge of a nearly Keplerian disc.  For a rapidly
rotating black hole the inner torus, which can be plausibly related to
coronal activity of the inner region, is located mainly within the
ergospheric region.

Observations of black hole systems  indicate that at high accretion
rates geometry of the innermost region is similar to that considered
in this paper.  High temperature of thermal component, dominating the
soft-state spectra of black hole binaries,  implies disc extending
down close to a black hole (e.g.\ Gierli\'nski \& Done 2004).  A
similar conclusion comes from the amount of X-rays reflected from the
disc in a Seyfert  galaxy, MCG--6-30-15 (Lee et al.\ 1999), which
seems to be  a high accretion rate object (Nowak \& Chiang 2000).
Furthermore,  Fe K$\alpha$ line profiles indicate a very central
location of the hard X-ray source in high accretion rate objects,
including XTE J1650-500 (Miller et al.\ 2002)  and GX 339-4 (Miller et
al.\ 2004) in their soft states as well as MCG--6-30-15 (Wilms et al.\
2001).   In our subsequent paper  we find that relativistic line
profiles observed in black hole systems  can be explained in a model
involving illumination of a disc by a central corona and that the
extreme profiles noted above  require such a   corona to be
constrained within the ergosphere.

Given the above observational evidence, we expect that results of this
paper should be relevant to modeling observations of objects    with
high luminosity to Eddington luminosity ratio.  On the other hand, we
emphasize   that nonthermal tails observed in some soft state spectra
(e.g.\ McConnell et al.\ 2002)  cannot be due to special or general
relativistic effects in a thermal plasma.  Such a tail most likely
indicates  that Comptonizing plasma contains nonthermal (in addition
to thermalized)  electrons  (e.g.\ Zdziarski 2000), which has not been
taken into account in this study.

Observations  in lower luminosity states, including most normal
Seyfert 1 galaxies and black hole binaries in their hard states,
require that the innermost part of accretion disc is replaced by an
optically thin flow,  with the optically thick material extending down
only to $20-50R_{\rm g}$ (e.g.\ Done, Madejski, \.Zycki 2000;  Miller
et al. 2001;  \.Zycki, Done \& Smith 1998; 1999).  Hard X-ray spectra
of such objects are well fitted by thermal  Comptonization models,
usually using a uniform temperature and density distribution, with
electron temperature within 50-100 keV and a Thomson optical depth
close to unity (e.g.\ Zdziarski 2000).  If the spacial extent of such
a uniform, hot plasma corresponds to the distance of truncation of the
disc,  then the relativistic effects considered in this paper should
not be important for the spectra.

However, if the plasma is heated due to dissipation of gravitational
energy  in a hot inner flow around a rotating black hole,  then
enhanced emission from the innermost regions can be expected.
Accretion flows around rapidly rotating black holes release a large
fraction of the gravitational power within the ergosphere, as noted
above.  Furthermore, advection increases electron temperature in the
innermost region of optically thin flows (Esin et al.\ 1997).  Then,
emission of an inner corona may be very centrally concentrated  and a
detailed modeling is required to check importance of  the
gravitational effects for formation of spectra in such a case.

Finally, we indicate that hints for orientation-dependent hard X-ray
spectra, consistent with predictions of our model, 
come from recent studies of X-ray
observations of Seyfert galaxies. In their analysis of OSSE
observations, Zdziarski, Poutanen \& Johnson (2000) found that Seyfert
2 galaxies have significantly harder spectra than Seyfert 1s, with the
average photon spectral index, in the 50 - 200 keV range, being
approximately 2 and 2.5, respectively. As only in one Seyfert 2,
considered in that paper, absorption of X-rays is Thomson thick, this
result may indicate an intrinsic difference between the hard X-ray
spectra of Seyfert 1s and 2s. A similar conclusion comes from {\it
BeppoSAX} observations (Deluit \& Courvoisier 2003), which indicate
also that the cut-off energy is higher in Seyfert 2s than in Seyfert
1s.

According to the unification model of active galactic nuclei (e.g.\
Antonucci 1993) the difference between type 1 and type 2 galaxies is
only due to the viewing angle. Namely, the same object would be
classified as a Seyfert 1 or as a Seyfert 2 when observed at low and
high inclinations, respectively. Following this unification scheme, we
indicate that  Seyfert 2 galaxies may indeed be intrinsically harder
than Seyfert 1s due to gravitational focusing of X-rays by rotating
black holes.

\section{Summary}

We have developed the Monte Carlo model for studying Comptonization
spectra emerging from  accretion flows in the Kerr metric.  We find
that strong gravity effects are crucial for the formation of the
spectra only in the ergosphere of a rotating black hole.  Spectra
involving a significant contribution from the ergosphere are strongly
dependent on the observer's inclination. Furthermore, intrinsic 
curvature and spectral breaks, related to strong gravity effects, 
may occur.

The high energy part of the ergospheric Comptonization spectrum is
significantly attenuated  when optically thick material is present in
the equatorial plane of the Kerr geometry.  Therefore, a noticeable
Comptonization radiation from the ergosphere is more likely in models
with a tenuous plasma occupying the innermost region.  

We indicate that hardening of Comptonization spectra  with increasing
inclination, which is predicted by Comptonization in the Kerr metric,
can be considered as an explanation for the difference between the
hard X-ray spectra of Seyfert 1 and Seyfert 2 galaxies.

\section*{Acknowledgments}
This research has been supported by grants from Polish Committee for
Scientific  Research (PBZ-KBN-054/P03/2001) and from \L \'od\'z
University. I am grateful to W\l odek Bednarek for his encouragement
and to Marek Sikora and prof.\ Maria Giller for their comments.
I also thank the referee for very careful reading of the manuscript 
and for valuable suggestions.

\label{lastpage}
\end{document}